\documentclass[prl,twocolumn,10pt, aps,superscriptaddress]{revtex4-2}

\usepackage{graphicx}
\usepackage{dcolumn}
\usepackage{bm}
\usepackage{graphicx}
\usepackage{color}

\begin{document}

\title{Carrier Localization and Spontaneous Formation of Two-Dimensional Polarization Domain in Halide Perovskites}

\author{Andrew Grieder}
\affiliation{Department of Materials Science and Engineering, University of Wisconsin, Madison Wisconsin 53718, USA}
\author{Marcos Calegari Andrade}
\affiliation{Department of Chemistry, University of California, Santa Cruz California 95064, USA}
\affiliation{Quantum Simulations Group, Lawrence Livermore National Laboratory, Livermore California 94551, USA}
\author{Hiroyuki Takenaka}
\affiliation{Department of Chemistry, University of California, Santa Cruz California 95064, USA}
\author{Tadashi Ogitsu}
\affiliation{Quantum Simulations Group, Lawrence Livermore National Laboratory, Livermore California 94551, USA}
\author{Liang Z. Tan}
 \email{lztan@lbl.gov}
\affiliation{Molecular Foundry, Lawrence Berkeley National Lab, Berkeley California 94720, USA}
\author{Yuan Ping}
 \email{yping3@wisc.edu}
\affiliation{Department of Materials Science and Engineering, University of Wisconsin-Madison, Madison Wisconsin 53718, USA}
\affiliation{Department of Physics, University of Wisconsin, Madison, Wisconsin, 53706, USA}
\affiliation{Department of Chemistry, University of Wisconsin, Madison, Wisconsin, 53706, USA}

\date{\today}

\begin{abstract}
Halide perovskites are known for their rich phase diagram and superior performance in diverse optoelectronics applications. The latter property is often attributed to the long electron-hole recombination time, whose underlying physical mechanism has been a long-standing controversy.
In this Letter, we investigate the transport and localization properties of electron and hole carriers in a prototypical halide perovskite (CsPbBr$_3$), through \textit{ab initio} tight-binding nonadiabatic dynamics approach for large-scale (tens of nm size) supercell calculations, to simulate electron and ion dynamics on the same footing. We found distinct structural, lattice polarization, and electron-phonon coupling properties at low (below 100 K) and high temperatures, consistent with experimental observations. In particular, at low temperature we find spontaneous formation of polar grain boundaries in the nonpolar bulk systems, which result in two-dimensional polarization patterns that serve to localize and separate electrons and holes. We reveal phonon-assisted variable-range hopping mostly responsible for low-temperature transport, and their characteristic frequency correlates with temperature-dependent phonon power spectrum and energy oscillation frequency in nonadiabatic dynamics.
We answer the critical questions of long electron-hole recombination lifetime at low temperature and offer the correlation among polarization domains, electron-phonon couplings, and photocarrier dynamics. 
\end{abstract}

\maketitle

Lead halide perovskites (LHPs) have received significant attention due to their remarkable optoelectronic properties, including low carrier recombination rates, high carrier mobility, and long carrier diffusion lengths~\cite{stranks_electron_hole_2013,herz_charge_carrier_2017,crothers_photon_2017,hutter_directindirect_2017}. The unique combination of properties has resulted in LHPs superior applications in solar cells, light-emitting diodes, and other electronic devices~\cite{snaith_perovskites_2013,kazim_perovskite_2014, stranks_metal_halide_2015}. Despite extensive research, the underlying mechanism remains under debate and thus demands clarification in order to widely apply LHPs in accessible technologies~\cite{Egger-Remaining-Issues,schilcher_significance_2021,martiradonna_riddles_2018}.

Proposed mechanisms to explain long electron-hole recombination lifetimes and diffusion lengths in LHPs include, charge carrier localization through formation of large polarons or ferroelectric domains. Large polarons form due to strong electron-phonon couplings and result in localization of excess holes or electrons coupled to lattice distortions over multiple unit cells~\cite{frost_calculating_2017, neukirch_polaron_2016,XYZ-largpolaron-jpclett-2015, XYZ-sciadv-largepolaron-2017,Wolf2017,miyata_lead_2017,Ambrosio2018}. However, it remains a subject of debate whether large polarons provide a complete explanation of carrier dynamics in LHPs~\cite{Sender-Frolich-RSC-2016,schilcher_significance_2021,wang_jacs_2021}. Alternatively, the formation of ferroelectric domains cause polar regions that force electron and hole trajectories to be physically separated and thus avoid recombination~\cite{Frost-Atomistic-Origins,Liu-Ferroelectric-Domain}. However, the presence or absence of ferroelectricity in LHPs has been a long-standing controversy. 

In this Letter, carrier dynamics and localization are investigated in a prototypical LHP, CsPbBr$_3$. The main challenges to accurately simulate carrier dynamics is the requirement to include a quantum mechanical description of electrons and holes, supercells on the order of tens of nanometers, and time propagation long enough to capture relevant dynamics. These requirements make direct quantum mechanical calculations intractable. To overcome the computational challenge, we adopt a combined technique utilizing a time-dependent first-principles tight-binding (FPTB) description for electrons and holes and machine learning (ML) interatomic potentials for ion-ion interactions. The dynamics are carried out under an Ehrenfest dynamics framework, including additional electronic forces from FPTB to account for the extra carriers in the system. By coupling ML force fields and tight-binding (TB) in this new framework, a quantum description of large supercells is possible and allows for discovery of phenomena missed by using ML or TB alone. This method directly addresses the issue of large-scale coupled electron-ion dynamics in a low symmetry structure, representing broad classes of problems such as electron-ion dynamics in ferroelectric relaxors, electron transfer, and dynamics at grain boundaries. Additionally, the framework could be adapted to study exciton dynamics in large complex systems by modifying the TB Hamiltonian to include additional electron-hole exchange interactions.

Accurate tight binding~\cite{boyer-richard2016symmetry-based,Zheng-EES-2019,nestoklon2021tight-binding}, and potential development~\cite{handley_new_2016,  hata_development_2017,gbalestra_efficient_2020}, including the use of ML~\cite{jinnouchi_phase_2019}, has been key to the study of halide perovskite dynamics. The FPTB model is constructed using physically motivated descriptors to efficiently and accurately compute the Hamiltonian given instantaneous ionic positions~\cite{abramovitch_thermal_2021}.  The FPTB and density functional theory (DFT) matrix elements agree with a root mean square error of  0.02 to 0.06 eV, significantly lower than thermal energy fluctuations (estimated to be $\sim$0.25 eV, see SI Section II).  The atomic positions are determined by machine learning (ML) inter-atomic potentials which allow for accurate descriptions of the octahedral tilting, not captured by standard pairwise potentials. The ML inter-atomic potentials in this study use the smooth version of the Deep Potential (DP) neural network architecture and reproduced DFT energies and forces with atomic force error less than 0.1eV/\AA , see SI section V and VI for more details~\cite{Zhang2019,Zhang2018,Zhang2018b}. Lastly, the charge carrier forces are computed from the instantaneous FPTB Hamiltonian in combination with the ML-forces to propagate the ions in time using LAMMPS ~\cite{LAMMPS}. Validation of all the components of the model can be found in the SI Sections II and VI. The combined approach allows for FPTB nonadiabatic dynamics with large supercells and sufficient statistical sampling to compute key observables and compare with past experimental and theoretical work~\cite{qian_photocarrier-induced_2023,wang_probing_2020,hutter_directindirect_2017}.

A key finding of this study is the natural formation of nano-domains at low temperature. During simulated annealing, polar 90$^\circ$ twinning grain-boundaries spontaneously form within the nonpolar bulk material. Such twinning grain-boundaries have also been observed in single crystal experiments of CsPbBr$_3$~\cite{twin-GB-exp,xie_effects_2024}, highlighting the accuracy of our theoretical method to predict inhomogeneous complex structure formation. The grain-boundaries induce planes of local lattice polarization along the grain-boundaries on the order of 20 $\mu C/cm^2$. The nature of the polarization patterns serves as a novel mechanism of electron-hole localization and separation.

Non-twinning grain-boundaries, have been proposed to increase recombination times ~\cite{Wang2019, xiao2020benign}. However, they involve tilting of entire crystalline domains and would not arise spontaneously, unlike the twin grain-boundaries considered here. In contrast, twinning  grain-boundaries do not contribute to carrier recombination in perovskites~\cite{xiao2020benign} and are even potentially beneficial as shown in previous atomistic studies based on static DFT calculations~\cite{Liu-Ferroelectric-Domain,zhang2023simple,warwick2019first}. This work offers significant improvement and additional insight by including a more comprehensive description of twinning domains, allowing for the direct observation of carrier localization and separation. The supercells in this work are 100 times larger  than pervious DFT-based studies, significantly reducing finite-size effects. Furthermore, we observe phonon-mediated hopping transport at the grain boundaries at low temperature, which cannot be calculated using standard first-principles approaches.

At high temperatures, the local polarization becomes dominated by thermal fluctuations,  and carrier transport transitions from hopping of localized carriers to band transport. The transition temperature between the two mechanisms is predicted by our mobility simulations and agrees well with experiments ~\cite{wang_probing_2020}. 

\begin{figure}
    \centering
    \includegraphics[width=\linewidth]{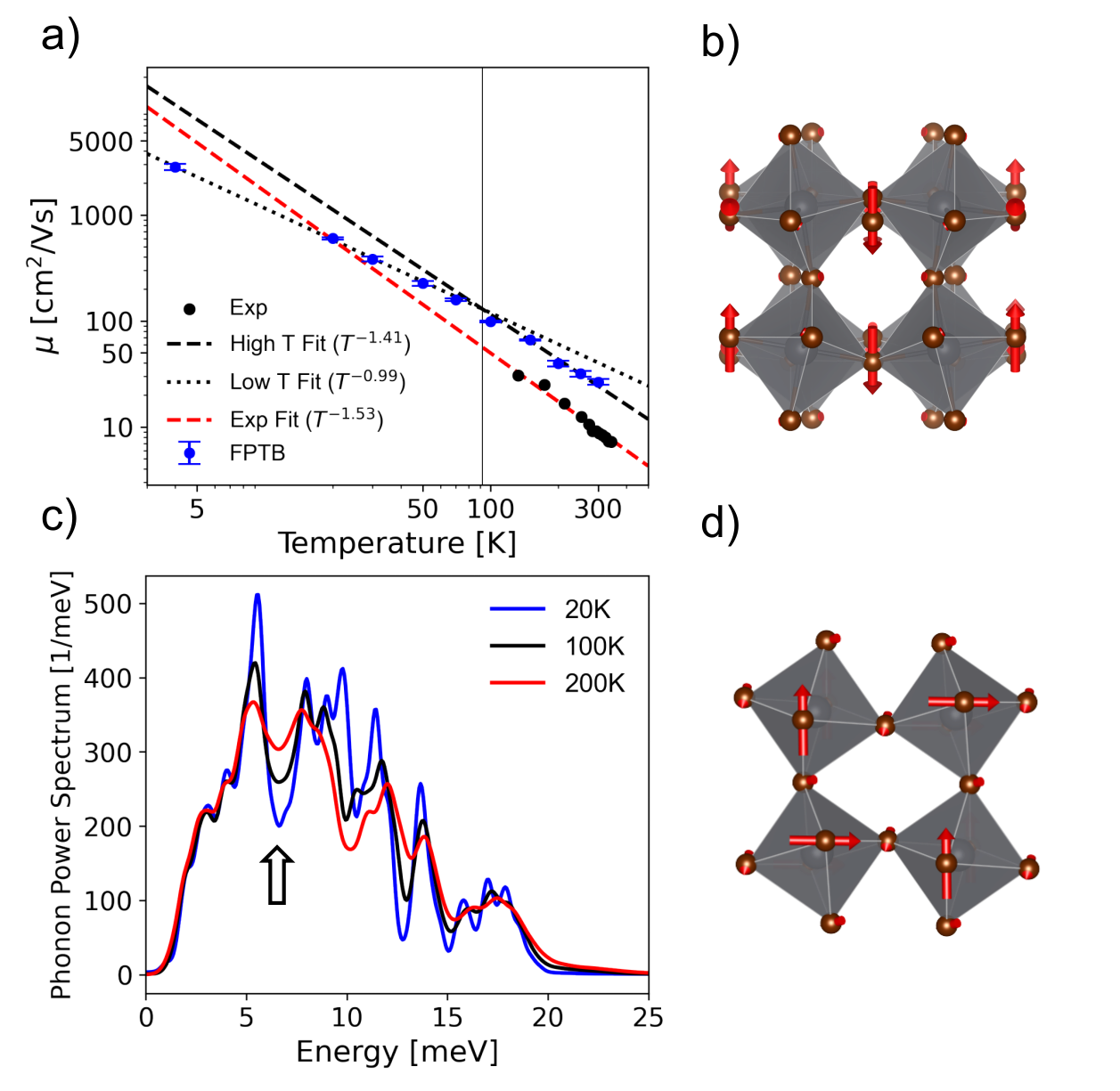}

    \caption{Computed temperature-dependent carrier mobility including important phonon contributions without nano-domains. a) Electron mobility at increasing temperature fit to low temperature (below 100 K) and high temperature range (above 100 K). Experiments (Exp) are from Ref \cite{hutter_directindirect_2017}. b) and d) Phonon modes whose occupations significantly increase with temperature at 7 meV, indicated in panel c). c) Phonon power spectrum of Br ions at 20, 100, and 200K.
    }
    \label{fig:mobility-phonon}
\end{figure}

\emph{Carrier mobility and temperature dependence}-- Electron mobility in CsPbBr$_3$ without nano-domains is computed for a range of temperatures to examine the electron-phonon coupling effects on transport properties. Diffusion coefficients ($D$) are determined from the spread of a Gaussian wave-packet in a 20x20x20 supercell (117.9x117.0x117.2 \AA). Carrier mobility ($\mu$) is then computed from diffusion coefficients using the Einstein relation ($\mu=eD(k_BT)^{-1}$) and reported in Fig.~\ref{fig:mobility-phonon}a), see SI Section III for details. The 20x20x20 supercell was found to be sufficiently converged with respect to supercell sizes, see SI Fig. S5

The computed mobility in the high-temperature regime $T>$100 K,  has a power law $T^{ -1.41}$ consistent with experiment \cite{wang_probing_2020,hutter_directindirect_2017}.  The magnitude of mobility is overestimated compared to experiments.  A temperature-independent defect scattering contribution is missing, which can lead to overestimation of mobility. Additionally, our TB model is fit to DFT-GGA with SOC, which is well known to underestimate the band gap and carrier effective masses compared to many-body methods such as GW approximation ~\cite{boyer-richard2016symmetry-based,jiang2025flexible}. The underestimation of carrier effective masses may lead to overestimation of carrier mobility as well.

To understand temperature dependent phonon contributions, the phonon mode projections and power-spectrum were analyzed at 20, 100, and 200 K (Fig.~\ref{fig:mobility-phonon}c). With increasing temperature, the Br projected phonon power spectrum shows significant increase in occupation around 7 meV and 13 meV and a decrease around 10 meV (Fig.~\ref{fig:mobility-phonon}c) . The modes responsible for the increase around 7 meV and 13 meV are identified to be the optical rocking modes as detailed in SI Section IV. Visualization of the optical rocking modes at 7 meV are presented in Fig.~\ref{fig:mobility-phonon}b and Fig.~\ref{fig:mobility-phonon}d , since they are important for low temperature transport presented below. The rocking octahedra cause the Pb-Br bond distances to fluctuate, impeding carrier mobility.  The mobility has different temperature power laws below and above $\sim$100 K, Fig.~\ref{fig:mobility-phonon}a, consistent with experiments\cite{wang_probing_2020}. Experimentally, the temperature dependence ($T^\beta$) of the mobility transitions at $\sim$ 80 K from $\beta > 0$ below 80 K to $\beta \approx -1.5$ above 80 K. The positive temperature dependence power law below 80 K indicates thermally-activated transport of localized carriers dominates at low temperature.

\begin{figure}[!ht]
    \centering
    \includegraphics[scale=0.5]{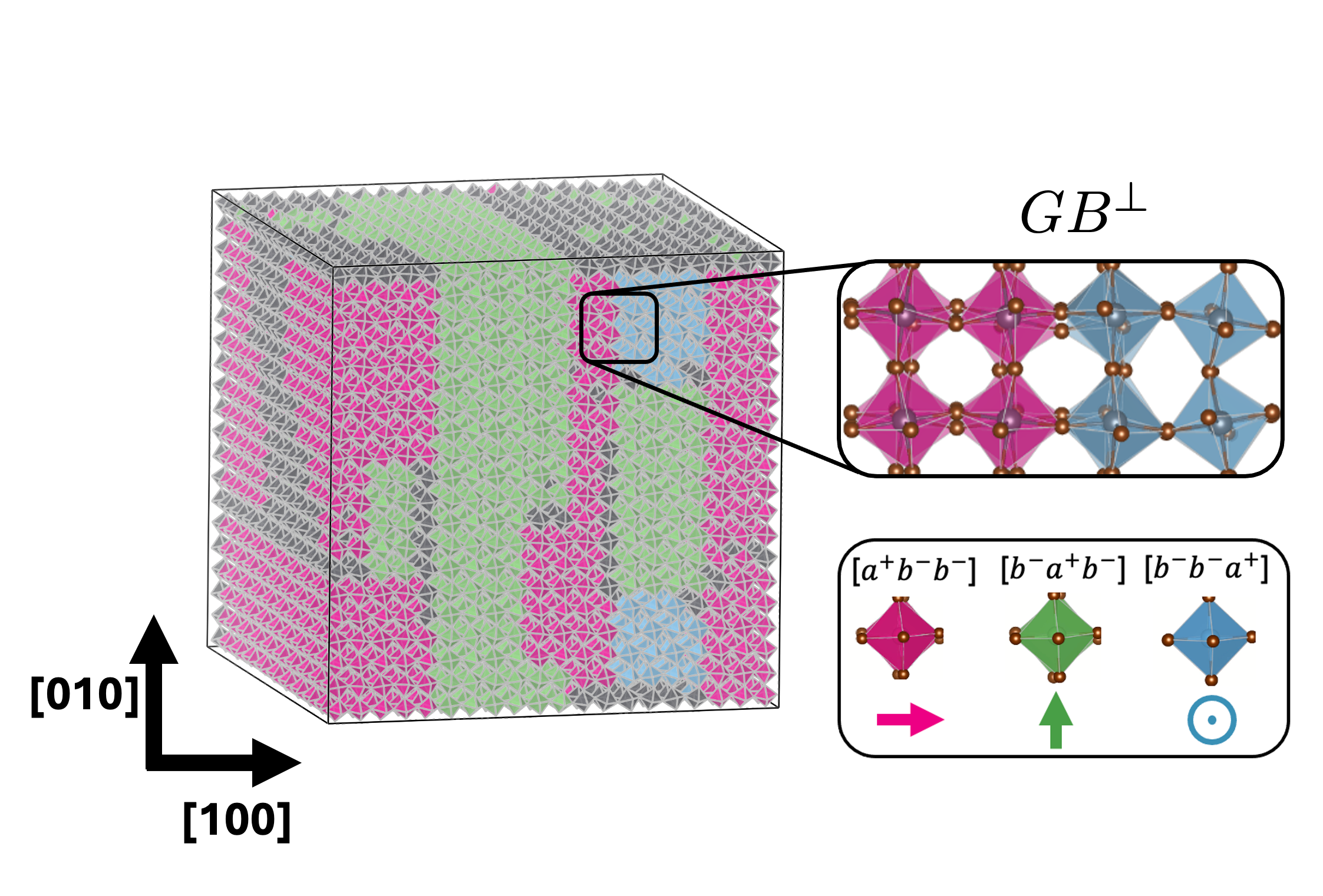}
    \caption{Grain-boundary formed during simulated annealing. Legends indicate grain-boundary classification and color coding for in-phase tilting axis, referred to as $ip$-axis in the text.}
    \label{fig:anneled-structure}
\end{figure}

\emph{Annealed structure and spontaneous formation of two-dimensional domains}-- The spontaneous formation of nano-domains occurs during the annealing simulation, see SI section VII. The nano-domains have twinning grain-boundaries with 90$^\circ$ rotations, consistent with experiments on single crystal CsPbBr$_3$ \cite{twin-GB-exp,xie_effects_2024}. Additionally, the grain-boundaries are found to be stable, lasting for longer than 100 ps of MD.

The individual domains are nonpolar, with $Pnma$ symmetry which has a Glazer notation of ($a^+b^-b^-$)~\cite{glazer_1972}. This structure has a single axis ($a^+$) with in-phase octahedral tilting and two axes ($b^-$) with out-of phase tilting. The unique in-phase tilting axis will be referred to as the $ip$-axis in the subsequent text. Excluding tilting of entire crystalline domains, there exist 3 possible orientations for the nano-domains, with $ip$-axis oriented along [100], [010], and [001], depicted in Fig.~\ref{fig:anneled-structure}.

The annealed structure  contains all three domain orientations with the $ip$-axis in the [100] (pink), [010] (green) direction, and [001] (blue) direction. In Fig.~\ref{fig:anneled-structure}, the domains can be seen to extend the entire (100) plane, indicating twinning grain-boundaries prefer parallel formation, consistent with experimental observation~\cite{twin-GB-exp,xie_effects_2024}. These 90$^\circ$ twinning grain-boundaries can be categorized into two types. The first type, denoted as $GB^{\perp}$, has domains with the $ip$-axis normal and parallel to the grain-boundary (Fig.~\ref{fig:anneled-structure}). In the second type, denoted as $GB^{\parallel}$, the $ip$-axis of both domains are parallel to the grain-boundary (SI Fig.~S10c). We observed formation of almost entirely $GB^{\perp}$suggesting the twinning boundaries have lower energy when the $ip$-axis is rotated perpendicularly rather than twisted. 

\emph{Carrier localization and variable range hopping}--  To investigate the effects of the domain structure on carrier dynamics, 
we simulate FPTB Ehrefenst dynamics with one extra electron or hole in the supercells with nano-domains. We note our simulations are at the mean-field level and do not include electron-hole recombination effects. However, we focus on carrier dynamics and find simulation times of 1.4 ps to be sufficient.
The initial wave function is set to the eigenstates of pristine CsPbBr$_3$ structure, using the VBM for holes and CBM for electrons. The initial condition ensures homogeneous distributions of the wave function, avoiding any bias in wave function localization. Note this is in contrast to the initial state in the mobility calculations where the wave function was initialized as a Gaussian wave-packet.

\begin{figure}
    \centering
    \includegraphics[width=\linewidth]{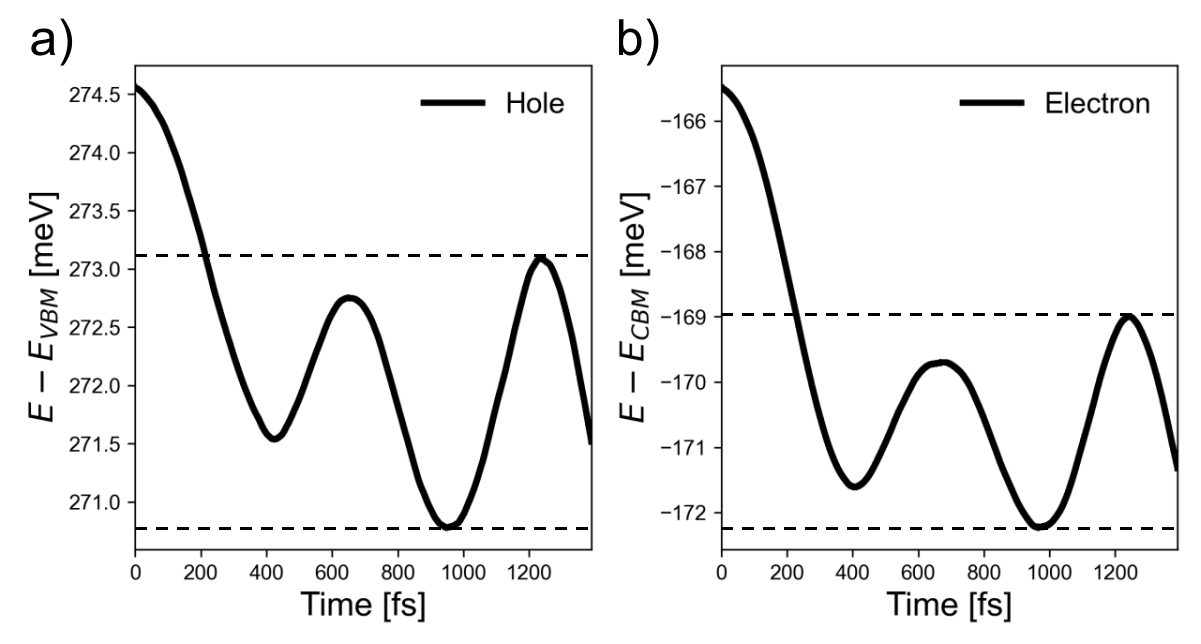}
    \caption{Change in energy from pristine eigenstate during Ehrenfest dynamics in the nano-domain structure. a) Holes with energy referenced to VBM and b) electrons with energy referenced to CBM of the pristine crystalline system.}
    \label{fig:electron-localization}
\end{figure}

The initial charge density is spread uniformly throughout the supercell and carriers become localized during the simulation. The hole and electron localize at spatially separated locations because of the local polarization as discussed below. Although the electron and hole are simulated separately, we expect minimal impact from electron-hole attraction. This is rationalized by the low exciton binding energy in CsPbBr$_3$ of 17-40 meV compared to the electron localization energy around 170meV~\cite{protesescu2015nanocrystals,shimosako2024exciton}. The electron localizes in a pancake shape as seen by the Gaussian fitting in SI Fig. S13, where the charge density is extended in the YZ-plane and narrow in the X direction. Both electrons and holes become localized after around 400 fs as seen in SI Fig. S12a. After localization, the energy oscillates around 170 meV below the CBM for electrons and 270 meV above the VBM for holes, Fig.~\ref{fig:electron-localization}. The oscillations in energy arise from interaction with phonons, indicating phonon-assisted electron hopping between localized sites. The phonon induced oscillations are confirmed in  SI Fig. S11 where the energy is constant in the absence of ionic motion.

\begin{figure*}[ht]
    \includegraphics[scale=0.8]{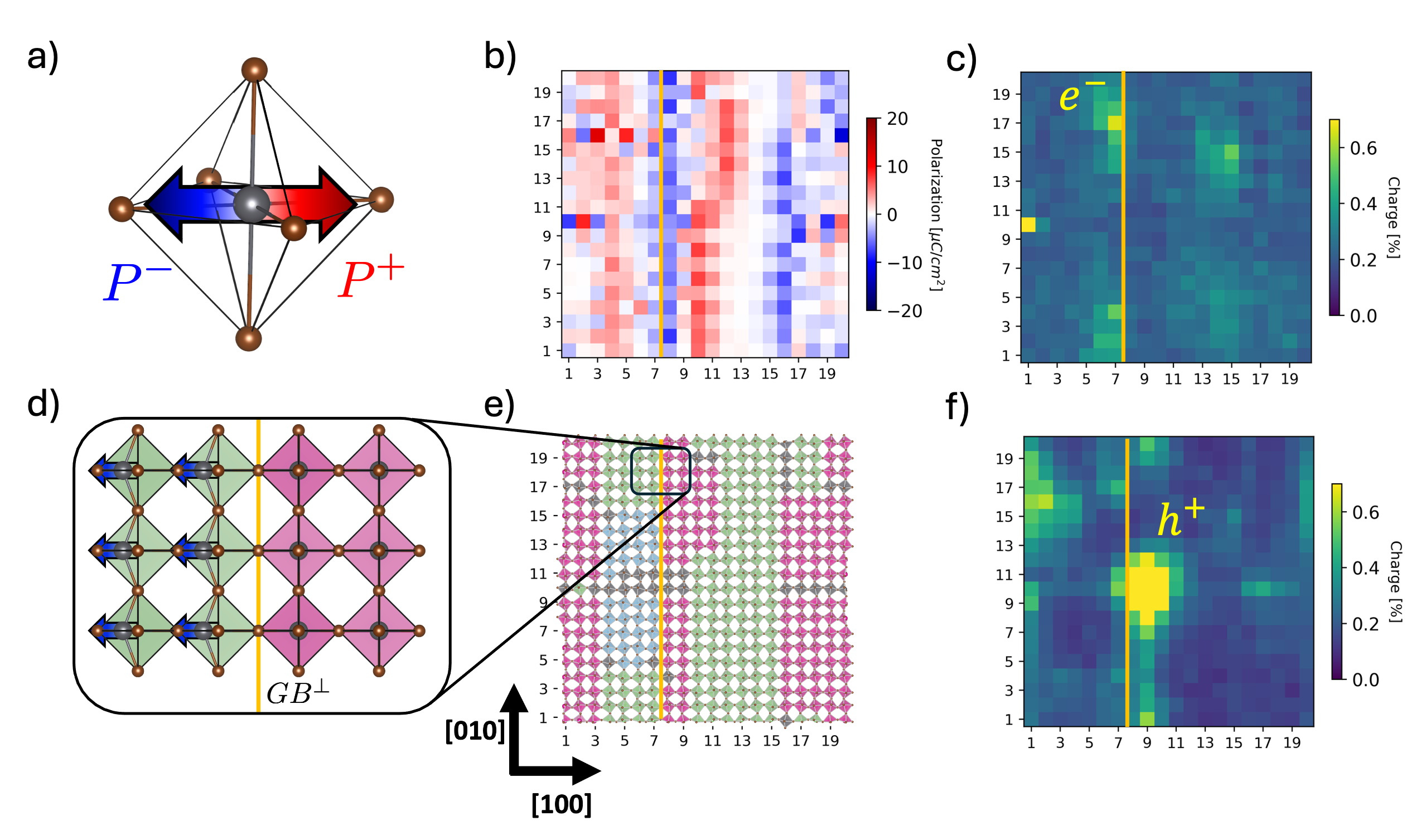}
    \caption {Relation among domain boundaries, local polarization, and electron/hole localization. a) Schematic depiction of local polarization ($P$) of $PbBr_6$ octahedra. b)Local polarization in the [100] direction averaged along the (001) plane. Polarization in units of $\mu C/cm^2$ . 2D projection of carrier density onto the (001) plane for electrons c) and holes f). The boundary between layer 7 and 8 is marked with an orange line in panels b), c), e), f). d) is schematic on how $GB^{\perp}$ induces dipole moment in layer with $ip$-axis parallel to the grain-boundary. e) Structure from simulated annealing with octahedra colored based on tilting pattern. }

    \label{fig:eh-seperation}
\end{figure*}

 The energy oscillation frequency for both the electron and hole in Fig.~\ref{fig:electron-localization} (around 1.7 THz) coincide with the critical optical phonon mode around 7 meV in Fig.~\ref{fig:mobility-phonon}c, further supporting the mechanism of phonon-assisted electron hopping.  The low temperature transport can be explained by the Mott's variable range hopping (VRH) given in Eq.~\ref{eq:mott-vrh}~\cite{mott1969conduction}. VRH requires disorder and a distribution of localized sites with various hopping distances, and has been previously used to study carrier transport in perovskites~\cite{bhaumik2024theoretical, moualhi2022fundamental}. In this work, the disorder of sites is caused by localization and disorder at the 2D polarization domains, thus we use the VRH to estimate low-temperature transport. Additionally, the VRH power law of  $T^{1/4}$, matches well with the experimental temperature-dependent mobility below 100 K~\cite{wang_probing_2020}:

\begin{equation}
 \label{eq:mott-vrh}
    \sigma = \sigma_0 \exp \left(- \left( \frac{T_M}{T} \right)^\frac{1}{4} \right)
 \end{equation}
 \begin{equation}
 \label{eq:sigma_prefactor}
     \sigma_0 = 3e^2 \nu_{ph} \left( \frac{g_0\xi}{8 \pi k_B T} \right)^{\frac{1}{2}}
 \end{equation}
 \begin{equation}
  \label{eq:t_m}
        T_M = \frac{\beta}{k_b g_0 \xi^3. } 
 \end{equation}

Here $e$ is the electron charge, $k_b$ is the Boltzmann constant, $\nu_{ph}$ is characteristic phonon frequency, $T$ is the temperature,  $\xi$ is the localization length, and $g_0$ is the density of states at the CBM such that the electron density at 20K is $2.0 \times 10^{15} cm^{-3}$  to match experimental carrier concentration~\cite{wang_probing_2020}.  The numerical constant $\beta$ is related to the critical percolation concentration, where we use $\beta = 21.2$ provided in Ref \cite{shklovskii2013electronic}. The localization length is estimated to be $\xi=1.2 \pm 0.1~nm$. It is computed from the full width half maximum of a Gaussian fitting of the electron charge density in the [100] direction averaged over the trajectory, with an example snapshot in Fig. S12. The characteristic phonon frequency is extracted from the energy oscillations to be $\nu_{ph}=1.7THz$. By using $\mu=\sigma/en $ with a carrier density of $2.0 \times 10^{15} cm^{-3}$, we obtain electron carrier mobility $\sim$ 4 $cm^2/Vs$. This is consistent with the experimental mobility measurements~\cite{wang_probing_2020} in dark in terms of order of magnitude (7$cm^2/Vs$) and temperature power-law ($T^{(1/4)}$)).

\emph{Local Polarization and Electron-Hole Separation} -- Finally, the local polarization of the PbBr$_6$ octahedra is computed to identify the effects of structural polarization on electron/hole localization (Fig.~\ref{fig:eh-seperation}). The local polarization is computed as $P=(q\cdot d)/V$ where $P$ is the polarization, $q$ is the Born Effective Charge tensor for Pb, $d$ is the displacement vector of the Pb ion from the center of the octahedra, and $V$ is the volume of the octahedra. This method for calculating local polarization of the PbBr$_6$ octahedra is consistent with the Berry-phase formalism, see SI Section VI.   
Ionic positions must be averaged over MD snapshots before computing local polarization to remove transient local polarization caused by thermal fluctuations. The net polarization was found to be minimized after 50 MD snapshots and thus used for all reported local polarizations. Although net polarization is close to zero after averaging (SI Fig. S9), the magnitude of local polarization remains at several tens of $\mu C/cm^2$ (Fig.~\ref{fig:eh-seperation}b), consistent with previous \textit{ab initio} calculations in CsPbBr$_3$ in the ferroelectric relaxor phase ~\cite{qian_photocarrier-induced_2023}. We note the local polarization is not uniform along the boundary and has a standard deviation of  $\sim$ 2 $\mu C/cm^2$.

The main grain boundary, $GB^{\perp}$, has no inversion or rotoinversion symmetry, thus allowing polarity. This manifests as Pb ions shifting away from the grain-boundary in the domain with $ip$-axis parallel to the boundary plane, penetrating about 3-4 layers deep, while the Pb ions in the other domain have minimal shift away from the center of octahedra. The overall dipole moment of $GB^{\perp}$ points normal to the grain boundary and is located in the domain with the $ip$-axis parallel to the boundary. This is illustrated in Fig.~\ref{fig:eh-seperation}d, and confirmed in the constructed grain-boundaries, SI Fig. S10b. The polarized domain always has the $ip$-axis parallel to the grain boundary, indicating the PbBr$_6$ octahedra are more easily polarized along the out-of-phase direction.

The lattice polarization induced by the grain-boundaries explains the nature of carrier localization. Being polar, $GB^{\perp}$ drives electrons to one side of the grain-boundary, while driving holes to the opposite side. This is supported by Fig.~\ref{fig:eh-seperation}c and f, which shows electrons driven in the direction of local dipole moment (i.e. in the same direction as the local displacement of Pb ions), and the holes driven in the opposite direction.  These 2D polarization planes  formed at the grain-boundaries provide a new mechanism for electron and hole localization and separation at low temperature. Although we did not observe this at room temperature, it may still be present in experiments. The damping of local polarization with increasing temperature may be enhanced by the limited size of our nano-domains compared to experiment. Due to the long lived nature of the grain-boundaries and local polarization, the electrons and holes tend to remain separated, thus resulting in long electron-hole recombination lifetime.

\emph{Conclusion}-- 
From our FPTB non-adiabatic dynamics simulations, a picture of hopping transport in CsPbBr$_3$ emerges at low temperature where electrons and holes are localized on the opposite sides of polar grain-boundaries of nano-domains. At low temperature, twinning grain-boundaries give rise to local polarization along the boundary. 
Given the experimental evidence of the 90$^\circ$ twinning grain-boundaries in single crystals, the domain boundaries and induced local polarizations offer a new mechanism to explain long electron-hole recombination lifetime in halide perovskites below 100 K. Because of the two-dimensional nature of nano-domain grain-boundaries, these results suggest a connection to experiments detecting reduced-dimensional transport in bulk halide perovskites. The possibility of such nano-domain grain-boundaries seeding or interacting with anisotropic electron-phonon effects such as 2D polarons or soft modes~\cite{lanigan-atkins_two-dimensional_2021,wang_jacs_2021,Ambrosio2018} remains to be investigated in the future. The FPTB non-adiabatic dynamics framework presented here directly addresses the issue of large-scale coupled electron-ion dynamics. This method is readily applicable to more complex perovskites such as mixed-halide or organic-inorganic and can be extended to other disordered systems.

\begin{acknowledgments}
\textbf{Acknowledgments} We acknowledge support from the Computational Materials Sciences Program funded by the US Department of
Energy, Office of Science, Basic Energy Sciences, Materials Sciences and Engineering Division for the materials application and development of \textit{ab initio} tight-binding code. 
We acknowledge support from the computational chemical science program within the Office of Science at
DOE under grant No. DE-SC0023301 for the code development of tight-binding-LAMMPS interfaces and parallelization. Additional support for data analysis and interpretation was provided by user program of the Molecular Foundry, Lawrence Berkeley National Laboratory, supported by the Office of Science, Office of Basic Energy Sciences, of the U.S. Department of Energy under Contract No. DE-AC02-05CH11231. Part of this work was performed under the auspices of the U.S. Department of Energy by Lawrence Livermore National Laboratory under Contract No. DE-AC52-07NA27344.
\end{acknowledgments}

\bibliography{ref}

\end{document}